\documentclass{elsart}
\usepackage[dvips]{graphicx}
\journal{Physics Letters A}
\begin{document}
\begin{frontmatter}

\title{Shock in a Branching-Coalescing Model with Reflecting Boundaries}
\author{Farhad H Jafarpour}
\address{Physics Department, Bu-Ali Sina University, Hamadan,Iran\\
Institute for Studies in Theoretical Physics and Mathematics
(IPM), P.O. Box 19395-5531, Tehran, Iran}
\ead{farhad@sadaf.basu.ac.ir}
\begin{abstract}
A one-dimensional branching-coalescing model is considered on a
chain of length $L$ with reflecting boundaries. We study the phase
transitions of this model in a canonical ensemble by using the
Yang-Lee description of the non-equilibrium phase transitions.
Numerical study of the canonical partition function zeros reveals
two second-order phase transitions in the system. Both transition
points are determined by the density of the particles on the
chain. In some regions the density profile of the particles has a
shock structure.
\end{abstract}

\end{frontmatter}
\begin{figure}[htbp]
\setlength{\unitlength}{1mm}
\begin{picture}(0,0)
\put(-4,26){\makebox{$\scriptstyle Im(q)$}}
\put(40,-2){\makebox{$\scriptstyle Re(q)$}}
\put(68,32){\makebox{$\searrow$}}\put(62,23){\makebox{$\nwarrow$}}
\put(65,36){\makebox{$q_c$}}\put(62,18){\makebox{$q_c'$}}
\end{picture}
\centering
\includegraphics[height=5cm] {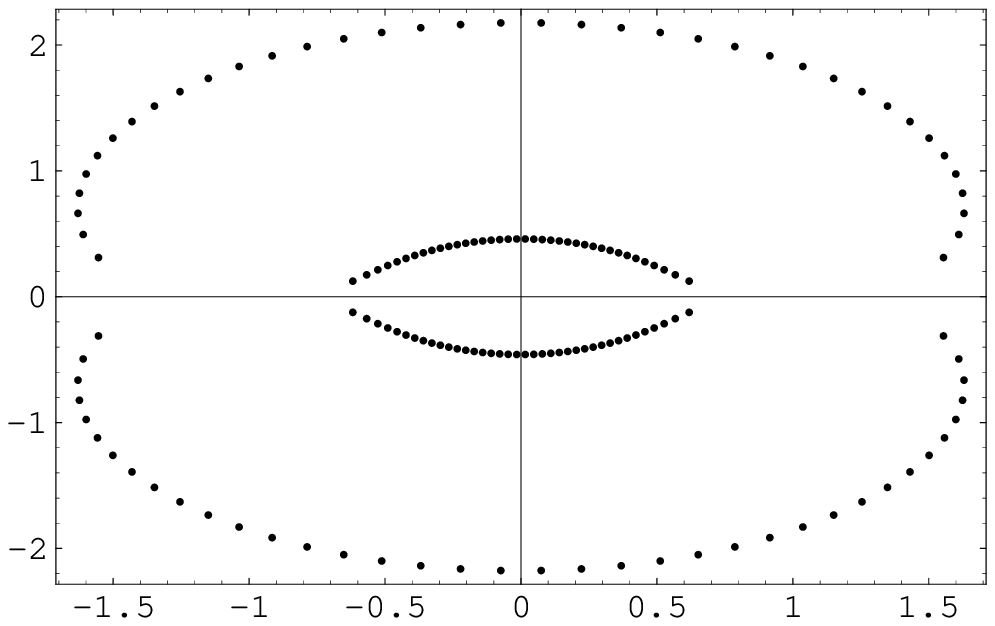}
\caption{\label{fig1} Plot of the numerical estimates for the
canonical partition function zeros obtained from (\ref{COEF1}) for
$L=80$ and $M=42$.}
\end{figure}
\begin{figure}[htbp]
\setlength{\unitlength}{1mm}
\begin{picture}(0,0)
\put(-4,26){\makebox{$\scriptstyle \rho(i)$}}
\put(45,-1){\makebox{$\scriptstyle i$}}
\end{picture}
\centering
\includegraphics[height=5cm] {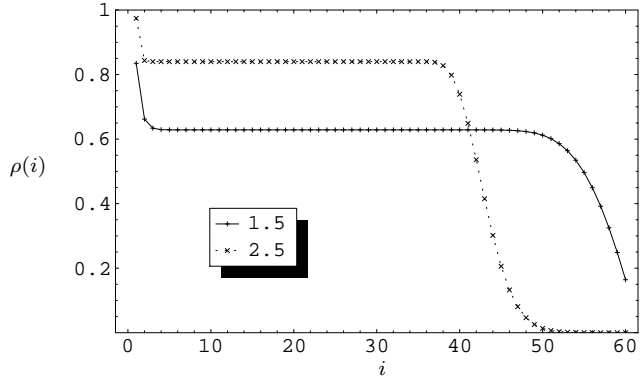}
\caption{\label{fig3} Plot of the density profile of the particles
(\ref{DP2}) on a chain of length $L=60$ for $M=36$ and two values
of $q \; (q>1)$ above and below the critical point $q_c=1.581$.}
\end{figure}

One-dimensional driven lattice gases are models of particles which
diffuse, merge and separate with certain probabilities on a
lattice with open, periodic or reflecting boundaries. In the open
boundaries case the particles are allowed to enter or leave the
system from both ends or only one end of the chain. In the
reflecting boundaries or periodic boundary cases the number of
particles will be a conserved quantity provided that no other
reactions other than the diffusion of particles take place. In the
stationary state, these models exhibit a variety of interesting
properties such as non-equilibrium phase transitions and
spontaneous symmetry breaking which cannot be found in equilibrium
models (see \cite{sch} and references therein). Different
applications are also found for such models which include the
kinetics of biopolymerization \cite{MGP} and traffic flow
modelling \cite{CSS}. These models have also allowed the study of
shocks i.e. discontinuities in the density of particles over a
microscopic region. Over the last decade people have studied the
shocks in one-dimensional driven-diffusive models with open and
periodic boundary conditions. A prominent example of such models
with open boundary is the Asymmetric Simple Exclusion Process
(ASEP) in which the particles enter the system from the left
boundary, diffuse in the bulk and leave the chain from the right
boundary with certain rates \cite{dehp}. For specific tuning of
injection and extraction rates a shock might appear in the system
which moves with a constant velocity towards the boundaries. The
ASEP has also been studied on a ring in the presence of a second
class particle called the impurity \cite{mall,lpk,farhad1}. In
this case the impurity will track the shock front with a constant
velocity which is determined by the reaction rates of the model.
The shocks in the models with reflecting boundaries have not been
studied yet.\\
In the present letter we study the phase transitions in a
one-dimensional branching-coalescing model with reflecting
boundaries in which the particles diffuse, coagulate and
decoagulate on a lattice of length $L$. The reaction rules are
specifically as follows:
\begin{equation}
\label{Rules}
\begin{array}{llll}
\mbox{Diffusion to the left:} && \emptyset+A \rightarrow
A+\emptyset &
\mbox{with rate} \; \; q \\
\mbox{Diffusion to the right:} && A+\emptyset \rightarrow
\emptyset+A &
\mbox{with rate} \; \; q^{-1} \\
\mbox{Coalescence to the left:} && A+A \rightarrow A+\emptyset &
\mbox{with rate} \; \; q \\
\mbox{Coalescence to the right:} && A+A \rightarrow \emptyset+A &
\mbox{with rate} \; \; q^{-1} \\
\mbox{Branching to the left:} && \emptyset+A \rightarrow A+A &
\mbox{with rate} \; \; \Delta q \\
\mbox{Branching to the right:} && A+\emptyset \rightarrow A+A
& \mbox{with rate} \; \; \Delta q^{-1}\\
\end{array}
\end{equation}
in which $A$ and $\emptyset$ stand for the presence of a particle
and a hole respectively. It is assumed that there is no injection
or extraction of particles from the boundaries. We will also
assume that the number of particles on the chain is a conserved
quantity. This model was first introduced and treated in continuum
approximation in \cite{db,bdb,bbd,dbh}. It was then studied using the
Empty Interval Method (EIM) in \cite{hkp1}. In this formalism the
physical quantities such as the density of particles are
calculated from the probabilities to find empty intervals of
arbitrary length. Later this model was studied using so-called the
Matrix Product Formalism (MPF) \cite{hkp2}. According to this
formalism the stationary probability distribution function of the
system is written in terms of the products of non-commuting
operators $E$ and $D$ and the vectors $\vert V \rangle$ and
$\langle W \vert$ as follows
\begin{equation}
\label{DF} P(\tau_1,\cdots,\tau_L)=\frac{1}{Z_L}\langle W \vert
\prod_{i=1}^{L} (\tau_iD+(1-\tau_i)E)\vert V \rangle.
\end{equation}
Each site of the lattice is occupied by a particle ($\tau_i=1$) or
is empty ($\tau_i=0$). The factor $Z_L$ in (\ref{DF}) is a
normalization factor. The operators $D$ and $E$ stand for the
presence of particles and holes respectively and besides the
vectors $\vert V \rangle$ and $\langle W \vert$ should satisfy the
following quadratic algebra \cite{hkp2}
\begin{equation}
\begin{array}{l}
\label{BulkAlgebra}
[E,\bar{E}] = 0 \\
\bar{E}D-E\bar{D}=q(1+\Delta)ED-\frac{1}{q}DE-\frac{1}{q}D^{2}\\
\bar{D}E-D\bar{E}=-qED+\frac{1+\Delta}{q}DE-qD^2\\
\bar{D}D-D\bar{D}=-q\Delta ED-\frac{\Delta}{q}DE+(q+\frac{1}{q})D^2\\
\langle W \vert\bar{E}=\langle W \vert \bar {D}=0 \\
\bar{E} \vert V \rangle =\bar{D}\vert V \rangle=0.
\end{array}
\end{equation}
The operators $\bar {D}$ and $\bar{E}$ are auxiliary operators and
do not enter into calculating (\ref{DF}). Having a representation
for the quadratic algebra (\ref{BulkAlgebra}) one can easily
compute the steady state weights of any configuration of the
system using (\ref{DF}). It has been shown that
(\ref{BulkAlgebra}) has a four-dimensional representation
\cite{hkp2}. For $q^2 \neq 1+\Delta$ we have
\begin{equation}
\label{RepBulk}
\begin{array}{c}
D=\left(\begin{array}{cccc}
0&0&0&0\\
0&\frac{\Delta}{1+\Delta}&\frac{\Delta}{1+\Delta}&0\\
0&0&\Delta&0\\
0&0&0&0
\end{array} \right),
E=\left(\begin{array}{cccc}
\frac{1}{q^2}&\frac{1}{q^2}&0&0\\
0&\frac{1}{1+\Delta}&\frac{1}{1+\Delta}&0\\
0&0&1&\frac{1}{q^2}\\
0&0&0&\frac{1}{q^2}
\end{array} \right),
\vert V \rangle=\left(\begin{array}{c} a\\0\\q^2\\q^2-1
\end{array} \right) \\ \langle W \vert=\Bigl( 1-q^2,1,0,b \Bigr)
\end{array}
\end{equation}
in which $a$ and $b$ are arbitrary constants. The matrix
representations for $\bar{D}$ and $\bar{E}$ are also given in
\cite{hkp2}. Both EIM and MPA approaches showed that the model has
two different phases: a low-density phase for $q^2>1+\Delta$ and a
high-density phase for $q^2<1+\Delta$ and a phase transition takes
place at the critical point where $q^2=1+\Delta$. In the
low-density and the hight-density regions the density profile of
the particles on the chain is an exponential function
while on the coexistence line it changes linearly along the lattice.\\
Recently this model has been studied under the open boundary
conditions on a specific manifold of the parameters of the system
\cite{farhad2}. It is shown that if the particles are injected and
extracted from the left boundary with the rates $\alpha$ and
$\beta$ respectively then the model has the same phase structure
provided that $\alpha=(q^{-1}-q+\beta)\Delta$. In the latter case
the operators $D$ and $E$ and also the vectors $\langle W \vert$
and $\vert V \rangle$ have two-dimensional representations. The
only difference is that for the reflecting boundary conditions the
system involves three different length scales while for the open
boundary conditions it is characterized by one length scale.
Moreover, it has been shown that in the open boundary case the
probability distribution function of the system can be written in
terms of superposition of Bernoulli shock measures \cite{kjs}. For
the open boundary case if we fix the density of particles, for
example by working in a canonical ensemble, we can see the real
shock structures in the density profile of particles. In this case
the system will also have two different phases: a low-density
phase and a jammed-phase where the shocks evolve in the system.
These phases are specified by the density of particles and are
separated by a second-order phase transition \cite{farhad3}.\\
A natural question that might arise is whether or not we can see
the shocks in our branching-coalescing model defined by
(\ref{Rules}) with the reflecting boundaries. To answer this
question we will study the model with reflecting boundaries in a
canonical ensemble where the number of particles on the chain is
equal to $M$ so that the density of particles $\rho=\frac{M}{L}$
remains constant. We will then investigate the phase transitions
and the density profile of particles on the chain. Recently it has
been shown that the classical Yang-Lee theory \cite{yanglee,gross}
can be applied to the one-dimensional out-of-equilibrium systems
in order to study the possible phase transitions of these models
\cite{be,arndt,farhad,farhad2,farhad3}. According to this theory
in the thermodynamic limit, the zeros of the canonical or grand
canonical partition function, as a function of an intensive
variable of the system, might approach the real positive axis of
that parameter at one or more points. Depending on how these zeros
approach the real positive axis the system might have one or more
phase transition of different orders. If the zeros intersect the
real positive axis at a critical point at an angle
$\frac{\pi}{2n}$, then $n$ will be the order of phase transition
at that point \cite{be}. \\ Let us define the canonical partition
of our model using the MPF as follows
\begin{equation}
\label{CPF} Z_{L,M}=  \sum_{\{\tau_i=0,1\}}
\delta(M-\sum_{i=1}^{L}\tau_i)\langle W \vert\prod_{i=1}^{L}
(\tau_i D+(1-\tau_i)E)\vert V \rangle
\end{equation}
in which $M$ and $L$ are the number of particles and the length of
the system respectively and $\delta(\cdots)$ is the ordinary
Kronecker delta function $\delta_{x,0}$. Using the matrix
representations $D$, $E$, $\langle W \vert$ and $\vert V \rangle$
given by (\ref{RepBulk}) we have been able to calculate the
canonical partition of this model (\ref{CPF}) and its zeros
numerically. One can use MATHEMATICA to calculate $\langle W \vert
(E+x D)^L\vert V \rangle$ for arbitrary $q$ and $\Delta$ and
finite $L$ in which $x$ is a free parameter. The result will be a
polynomial of $x$. The coefficient $x^M$ in this polynomial gives
the canonical partition function of the system. Formally we can
write
\begin{equation}
\label{COEF1} Z_{L,M}=Coefficient[\langle W\vert (E+xD)^L\vert V
\rangle,M]
\end{equation}
in which $Coefficient[Expr,n]$ gives the coefficient of $x^n$ in
the polynomial $Expr$. In Fig.~\ref{fig1} we have plotted the
numerical estimates for the zeros of $Z_{L,M}$ obtained from
(\ref{COEF1}) on the complex-$q$ plane for $L=80$, $M=42$. The
canonical partition function (\ref{COEF1}) has $4(L-M)$ zeros in
the complex-$q$ plane. We have found that for large $L$ and $M$
the locations of these zeros are not sensitive to the value of
$\Delta$. We have also calculated the numerical estimates for the
roots of (\ref{COEF1}) as a function of $\Delta$ for fixed values
of $q$. It turns out that (\ref{COEF1}), as a function of
$\Delta$, does not have any positive root; therefore, we expect
that the phase transition points do not depend on $\Delta$.
As can be seen in Fig.~\ref{fig1} the zeros lie on two different
curves and accumulate towards two different points on the positive
real-$q$ axis. By extrapolating the real part of the nearest roots
to the positive real-$q$ axis for large $L$ and $M$, we have found
that the transition points are $q_c=\frac{1}{\sqrt{1-\rho}}\;\;(1
< q_c < \infty)$ and $q_c'=\sqrt{1-\rho}\;\;(0 < q_c' < 1)$. As
$\rho \rightarrow 0$ the two curves lie on each other and we will
find only one transition point at $q_c=q_c'=1$. It appears also
that the zeros on both curves approach the real-$q$ axis at an
angle $\frac{\pi}{4}$ (the smaller angle). This predicts two
second-order phase transitions at $q_c$ and $q_c'$. The reason
that the system has two phase transitions can easily be
understood. The parameter $q$ determines the asymmetry of the
system and for any $q$ the system is invariant under the following
transformations
\begin{eqnarray}
\label{Symmetry}
q&\longrightarrow &q^{-1} \nonumber \\
i&\longrightarrow &L-i+1.
\end{eqnarray}
Therefore, one can expect to distinguish two critical points which
are related according to the symmetry of the system.\\
Let us now study the density profile of particles on the chain
$\rho(i)$ in each phase. The density of particles at site $i$ is
defined as
\begin{equation}
\label{DP1} \rho(i)=\frac{\sum_{{\mathcal C}}\tau_i
P(\tau_1,\cdots,\tau_L)}{\sum_{{\mathcal
C}}P(\tau_1,\cdots,\tau_L)}
\end{equation}
in which ${\mathcal C}$ is any configuration of the system with
fixed number of particles $M$ and $P(\tau_1,\cdots,\tau_L)$ is
given by (\ref{DF}). It can be verified that the density profile
of the particles $\rho(i)$ can be written as
\begin{equation}
\begin{array}{c}
\label{DP2}  \rho(i)=\frac{1}{Z_{L,M}}\\ \sum_{k=0}^{M-1} \langle
W \vert Coefficient[C^{i-1},k] \; D \;
Coefficient[C^{L-i},M-k-1]\vert V \rangle
\end{array}
\end{equation}
where we have defined $C:=E+xD$. Now one can use the matrix
representation (\ref{RepBulk}) to calculate (\ref{DP2}) using
MATHEMATICA. In Fig.~\ref{fig3} we have plotted (\ref{DP2}) for
two different values of $q$ with $L=60$ and $M=36$.
For this choice of the parameters the transition points are
$q_c=1.581$ and $q_c'=0.633$. The density of particle has two
general behaviors for $q > 1$. For $q > q_c$ and in the
thermodynamic limit
($L\rightarrow\infty,M\rightarrow\infty,\rho=\frac{M}{L}$) the
density profile of particles is a shock in the bulk of the chain;
while in the close vicinity of the left boundary, it increases
exponentially. The density of particles in the hight-density
region of the shock is equal to $\rho_{High}=1-q^{-2}$. This
region is separated by a rather sharp interface from the
low-density region in which the density of particles is equal to
$\rho_{Low}=0$. The low-density region is extended over
$(1-\frac{\rho}{1-q^{-2}})L$ sites. This can be seen in
Fig.~\ref{fig3} for $q=2.5$; however, the reason that the shock
interface is not sharp is that our calculations are not done in
real thermodynamic limit. One should expect that the shock front
becomes sharper and sharper as the length of the system $L$ and
also the number of particles on the chain $M$ increase. For $1 < q
< q_c$ the density of particles in the bulk of the chain is
constant equal to $\rho$, it drops near the right boundary
exponentially and increases exponentially in the close vicinity of
the left boundary. The exponential behavior of the density profile
of particles near the boundaries in this phase is due to the
finiteness of the representation of the algebra
(\ref{BulkAlgebra}). It is known that if the associated quadratic
algebra of the model has finite dimensional representations, the
density-density correlation functions cannot have algebraic
behaviors \cite{hkp2}. At $q=1$ one finds $\rho(i)=\rho$. The
density profile of particles in the region $q < 1$ is related to
that of $q > 1$ through (\ref{Symmetry}) that is
\begin{equation}
\label{DP3} \rho(i,q)=\rho(L-i+1,q^{-1}).
\end{equation}
One can also study this model on a ring of length $L$ with
periodic boundary conditions. Let us assume that the number of
particles and therefore their density fluctuates and is not a
constant. In this case the probability of finding the system in a
specific configuration ${\mathcal C}=\{\tau_1,\cdots,\tau_L\}$
should be obtained from
\begin{equation}
\label{DFR} P(\tau_1,\cdots,\tau_L)=\frac{1}{Z_L}
tr[\prod_{i=1}^{L} (\tau_iD+(1-\tau_i)E)].
\end{equation}
in which $tr[\cdots]$ is the trace of the products of matrices.
The normalization factor $Z_L$ can be obtained from the fact that
$\sum_{\mathcal C}P({\mathcal C})=1$. This function plays the role
of the grand canonical partition function of the system. One can
easily check that the quadratic algebra associated to the periodic
boundary condition case is the same as (\ref{BulkAlgebra}) except
the boundary terms which contain the vectors $\vert V \rangle$ and
$\langle W \vert$; therefore, we can still use the algebra
(\ref{BulkAlgebra}) and its representation (\ref{RepBulk}) to
calculate (\ref{DFR}). As we mentioned the grand canonical
partition function of the system can be defined as
\begin{equation}
\label{GCPF1} Z_{L}=\sum_{\{ \tau=0,1 \}}tr[\prod_{i=1}^{L}
(\tau_iD+(1-\tau_i)E)]=tr[(E+D)^L].
\end{equation}
This can easily be calculated and we find
\begin{equation}
\label{GCPF2} Z_{L}=1+q^{-2L}+q^{2L}+(1+\Delta)^L.
\end{equation}
The study of the zeros of (\ref{GCPF2}) as a function of $q$ in
the thermodynamic limit $L\rightarrow \infty$ shows that they
approach the positive real-$q$ axis at two different points
$q_c=\sqrt{1+\Delta} \; (1 < q_c < \infty)$ and
$q_c'=\frac{1}{\sqrt{1+\Delta}} \; (0 < q_c' < 1)$. Moreover, the
zeros approach the positive real-$q$ axis at angle
$\frac{\pi}{2}$; therefore, unlike the reflecting boundaries case
both phase transitions in this case are of first-order. In the
steady state the density profile of the particles on the ring is
flat and equal to $\rho(i)=\frac{\Delta}{1+\Delta}$ for $q_c' < q
< q_c$ and $\rho(i)=0$ for $q > q_c$ and $q < q_c'$. One can take
the total density of particles on the ring as an order parameter
and since it changes discontinuously over the transition points,
as the Yang-Lee theory predicts, the phase transitions are of
first-order. One should note that in comparison to the reflecting
boundaries case not only the geometry of the model is changed but
also the number of particles in this case in not a conserved
quantity. This means that we should not expect the transitions in
these models to be similar even if we take the thermodynamic limit
$L\rightarrow\infty$. The reason for the existence of two phase
transition points is again the
symmetry of the model. \\
In this letter we have studied a branching-coalescing model in
which particles hop, coagulate and decoagulate on one-dimensional
lattice of length $L$. We have restricted ourself to the case
where the total number of particles on the chain is constant. For
this we have worked in the canonical ensemble in which the number
of particles $M$ is fixed; therefore, the density of particles on
the chain $\rho=\frac{M}{L}$ is constant. The Yang-Lee theory
predicts that the model has two second-order phase transitions.
Both phase transition points are determined by the density of
particles on the system $\rho$. The study of the mean particle
concentration at each site of the chain for $q > 1$ shows that the
density profile of the particles has a shock-like structure in the
region $q > q_c=\frac{1}{\sqrt{1-\rho}}$. The exception is near
the left boundary where the density of particles increases
exponentially. This is the first time that shocks are seen in
one-dimensional reaction-diffusion models with reflecting
boundaries. In the region $1 < q < q_c=\frac{1}{\sqrt{1-\rho}}$
the density profile of the particles is constant in the bulk of
the chain; however, near the left (right) boundary it increases
(decreases) exponentially. The exact form of the correlation
lengths is under investigation. Our numerical investigations also
show that the width of the shock scales as $L^{-\nu}$ with
$\nu=\frac{1}{2}$. In the thermodynamic limit $L\rightarrow\infty$
the shock width goes to zero and one finds a very sharp shock interface.
Since the system is invariant under the
transformation (\ref{Symmetry}), the density profile of the
particles for $q < 1$ can be obtained from (\ref{DP3}). We also
studied the periodic boundary case and found that the system
possess two first-order phase transitions which are determined by
the values of $\Delta$ for fixed value of $q$. In this case the
density profile of particles is flat everywhere on the lattice and
is either equal to $\frac{\Delta}{1+\Delta}$ or zero. The formulas
(\ref{COEF1}) and (\ref{DP2}) provide us with a simple and general
tool for numerical study of the phase transitions and also the
particle concentration behaviors of one-dimensional stochastic
models in canonical ensemble for which a finite- or
infinite-dimensional representation of the associate algebra
exists.

\end{document}